\documentclass[11pt]{article}
\usepackage{moriond,epsfig}

\def\Journal#1#2#3#4{{#1} {\bf #2}, #3 (#4)}

\def\mnras{{\em MNRAS}}
\def\apj{{\em ApJ}}

\def\aap{{\em A\&A}}
\def\aj{{\em AJ}}
\def\nat{{\em Nature}}

\def\lesssim{\mathrel{\hbox{\rlap{\hbox{\lower4pt\hbox{$\sim$}}}\hbox{$<$}}}}
\def\gtrsim{\mathrel{\hbox{\rlap{\hbox{\lower4pt\hbox{$\sim$}}}\hbox{$>$}}}}
\def\phn{\phantom{0}}
\def\phd{\phantom{.}}
\def\phs{\phantom{$-$}}
\newcommand{\etal}{{\it et~al}}
\newcommand{\Fig}[1] {Fig.~\ref{#1}}
\newcommand{\Tab}[1] {Table~\ref{#1}}
\newcommand{\scinum}[2]{#1\times 10^{#2}}
\newcommand{\hmpc}[1]{\mbox{$#1 h^{-1} \mbox{Mpc}$}}
\newcommand{\hkpc}[1]{\mbox{$#1 h^{-1} \mbox{kpc}$}}
\newcommand{\cobe}{\textit{COBE}}
\newcommand{\tred}{three-dimensional}
\newcommand{\nbody}{$N$-body}
\newcommand{\pppm}{$\mathrm{P}^3\mathrm{M}$}
\newcommand{\Msol}{M_\odot}
\newcommand{\hMsol}{h^{-1} \Msol}
\newcommand{\OhMsol}{\Omega_0 \, \hMsol}
\newcommand{\Lsol}{L_\odot}
\newcommand{\Lstar}{$L_\star$}
\newcommand{\Mstar}{\mbox{$M_\star$}}
\newcommand{\OmegaB}{\Omega_{\mathrm{B0}}}
\newcommand{\cc}{\lambda_0} 
\newcommand{\Lbox}{L_{\rm box}}
\newcommand{\HUnits}{\,\mathrm{km}\,\mathrm{s}^{-1} \mathrm{Mpc}^{-1}}

\begin{document}
\vspace*{4cm}
\title{\cobe- AND CLUSTER-NORMALIZED CDM SIMULATIONS}

\author{ T.~PIRAN$^\dag$, H.~EL-AD$^{\dag \ast}$, H.~MARTEL$^\S$ and
  M.~LECAR$^\ast$ }

\address{
  $^\dag$Racah Institute for Physics, The Hebrew University, Jerusalem, 91904 Israel\\
  $^\ast$Harvard-Smithsonian Center for Astrophysics, 60 Garden Street, Cambridge, MA 02138, USA\\
  $^\S$Department of Astronomy, University of Texas, Austin, TX 78712, USA}

\maketitle\abstracts{ We introduce a set of four new publicly
  available \nbody{} simulations, the most recent additions to the
  \textit{Texas \pppm{} Database}. Our models probe the less studied
  parameter space region of moderate volume (\hmpc{100} box) combined
  with fine mass resolution ($\propto 10^{12} \Msol$, roughly
  comparable to a \Lstar{} galaxy), making these simulations
  especially suitable for study of major large-scale structure (LSS)
  features such as voids, and for comparison with the largest \tred{}
  redshift surveys currently available.  Our cosmological models
  (LCDM, TOCDM, OCDM, TCDM) are all \cobe-normalized, and when
  possible (LCDM and TOCDM) also cluster-normalized, based on the
  X-ray cluster $M$--$T$ relation.  The \cobe- and cluster-normalized
  LCDM model reiterates the attractiveness of this currently favored
  model which does not require the introduction of tilt in order to
  fit the constraints imposed by observations of other cosmological
  parameters.  }

\section{Introduction}
\label{intro}

\nbody{} simulations are an essential tool for probing LSS and galaxy
formation. As large, high-resolution simulations are computationally
costly, one has to carefully consider the added effort resulting from
increasing the simulations' volume or from improving their resolution.
In this context, most simulations gravitate towards a design stressing
either of these two conflicting goals. The largest \tred{} redshift
surveys currently available are situated somewhere in between these
two extremes: a $ \Mstar $ galaxy in the CfA2 survey is visible out to
\hmpc{100}. A simulation designed to match these surveys must have
both the required resolution to identify the dark matter (DM) halos
associated with such galaxies, \emph{and} this moderately large
volume.

The other essential consideration of simulation design is the choice
of cosmological models probed. Ideally, one would want to examine a
certain range of the relevant cosmological parameters ($H_0$,
$\Omega_0$, $\cc$, $\OmegaB$, $n$, $\sigma_8$), but this is often not
an attainable goal. In this work we focus on cosmological models with
currently favored values of $H_0$, $\Omega_0$ (and $\cc$), and require
that all models will be \cobe-normalized.
The above constraints can still be fulfilled through a variety of
primordial power spectrum tilt $n$ and $\sigma_8$ combinations. We
attempt to achieve also cluster normalization, hence determining the
value of $\sigma_8$ (and thus fixing a tilt value).

\section{Models}
\label{models}

These simulations were designed with the goal of maximizing their
volume while still being able to resolve DM halos associated with
\Lstar{} galaxies. Adopting a mass-to-light ratio of $100
\Msol/\Lsol$, we thus require that we will be able to identify $
M_{\mathrm{halo}} \gtrsim 10^{12} \Msol$ halos.  A simulation with
$140^3$ particles in a \hmpc{100} box will have $
M_{\mathrm{particle}} = \scinum{1.01}{11} \OhMsol $. Identifying all
DM halos with 20 particles or more, this parameter specification
matches the stated requirements.

\begin{table}[t]
\caption{Simulation Parameters}
\vspace{0.3cm}
\begin{center}
\begin{tabular}{|c|c|c|c|c|}
\hline
$\Lbox$ & $N_{\rm cell}$ & $L_{\rm cell}$ & $N_{\rm particles}$ & $M_{\rm particle}$ \\
\mbox{[\hmpc{}]}&         &[\hkpc{}]&         & [$\OhMsol$] \\
\hline
    100         & $256^3$ & 390     & $140^3$ & $\scinum{1.01}{11}$ \\
\phn 40         & $256^3$ & 156     & $120^3$ & $\scinum{1.02}{10}$ \\
\hline
\end{tabular}
\end{center}
\label{param-tab}
\end{table}

Failure to identify halos all the way down to this resolution renders
such simulations inappropriate for the study of many LSS features.
Specifically, it would be impossible to identify correctly voids in
such simulations.  Such studies often originate from DM
simulations, and use some form of halo populating scheme in order to
match the observed properties of the distribution of galaxies.  These
populating schemes (also known as bias recipes) assign a {\em number}
of galaxies to each halo. But for such schemes to have a fighting
chance at reproducing the distribution of galaxies, one must initially
know the {\em locations} of the halos that should be
populated---including halos that would be populated by just one
galaxy.

We estimated that 20 particles per halo is the minimal number required
in order to reliably identify halos. As this is getting close to the
simulation's mass resolution limit, we tested the lower end of our DM
halo mass function by constructing a matching set of smaller
simulation boxes with $M_{\mathrm{particle}}$ an order of magnitude
smaller~\cite{gross98} (see \Tab{param-tab}). We can then compare the
number density of DM halos in the two sets of boxes and see if indeed
we manage to recover the correct number of small halos in the larger
simulation boxes.

All simulations were started at an initial redshift $z_i = 24$ and
evolved over 600--1000 timesteps using a \pppm{} code over a $256^3$
grid. More details on the simulations can be found
elsewhere.~\cite{mm00,mme00}

In \Tab{cosmo-tab} we summarize the cosmological parameters defining
our models. Column~1 indicates the models' acronyms. Columns 2--5
detail the models' values of the Hubble parameter $h$, density
parameter $\Omega_0$, cosmological constant $\cc$, and tilt $n$.
Column~6 lists the theoretical $\sigma_8^{\rm cluster}$ based on X-ray
cluster temperatures. Column~7 lists $\sigma_8^{\rm cont}$, the actual
value corresponding to each cosmological model.  If these two values
match, we state that the model is cluster-normalized (Column~8).

\begin{table}
\caption{Model Parameters}
\vspace{0.3cm}
\begin{center}
\begin{tabular}{|l|c|c|c|c|c|c|c|c|}
\hline
Model & $h$ & $\Omega_0$ & $\cc$ & $n$ & $\sigma_8^{\rm cluster}$ & $\sigma_8^{\rm cont}$ & Cluster-Normalized? \\
\phs \phd(1) & (2) & (3) & (4) & (5) & (6) & (7) & (8) \\
\hline
TOCDM& 0.65 & 0.3 & 0\phd\phn & 1.3 &$0.91\pm0.09$& 0.91 & yes\\
OCDM & 0.65 & 0.3 & 0\phd\phn &1\phd\phn&$0.91\pm0.09$& 0.46 & no\\
LCDM & 0.65 & 0.3 &0.7&1\phd\phn&$1.00\pm0.09$& 0.95& yes\\
TCDM & 0.55 &1\phd\phn &0\phd\phn &0.7 &$0.53\pm0.05$& 0.72& no \\
SCDM & 0.5\phn&1\phd\phn&0\phd\phn&1\phd\phn&$0.53\pm0.05$& 1.27 & no \\
\hline
\end{tabular}
\end{center}
\label{cosmo-tab}
\end{table}

We fixed all models to be \cobe-normalized with $T_{\mathrm{CMB}} =
2.7 K$, assuming no contribution from tensor modes.  Also, we used $
\OmegaB = 0.015 h^{-2} $ throughout,~\cite{copi95} in concordance with
the \textit{Texas \pppm{} Database}.  When practical (all models but
TCDM) we adopt a Hubble constant~\cite{riess98} $H_0 = 65 \HUnits$.
In addition to \cobe{} normalization, we have also tried to achieve
cluster normalization.  Using the X-ray cluster $M$--$T$
relation~\cite{pen98} we derived the required $\sigma_8$ value for
each of our models. For each combination of $\Omega_0$, $\cc$, and
$H_0$, we computed the tilt required in order to achieve cluster
normalization and examined whether it is acceptable in view of
the limits allowed by the 4-year \cobe{} data.

For two of the models---LCDM and TOCDM---we found an acceptable tilt
value and managed to achieve cluster normalization.  The two other
models are not cluster-normalized. The OCDM model was designed as a
direct companion to the LCDM model, where all the cosmological
parameters in both these models---except the value of $\cc$---are the
same.  The TCDM model is our best attempt with an $\Omega_0 = 1$
model, where we used the lowest possible $\sigma_8$ value which does
not require $h < 0.55$ or $n < 0.7$. For comparison, we have also
included in \Tab{cosmo-tab} (and in \Fig{cmb-fig}) the familiar SCDM
model, although it is not one of the cosmological models simulated.

\section{Results}
\label{results}

\begin{figure}
\begin{center}
\psfig{figure=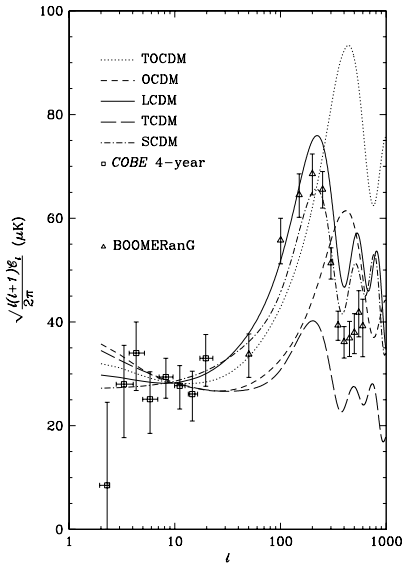,height=2.95in}
\caption{CMB Angular Power Spectra: Observations vs.\ Models}
\label{cmb-fig}
\end{center}
\end{figure}

\begin{figure}
\begin{center}
\psfig{figure=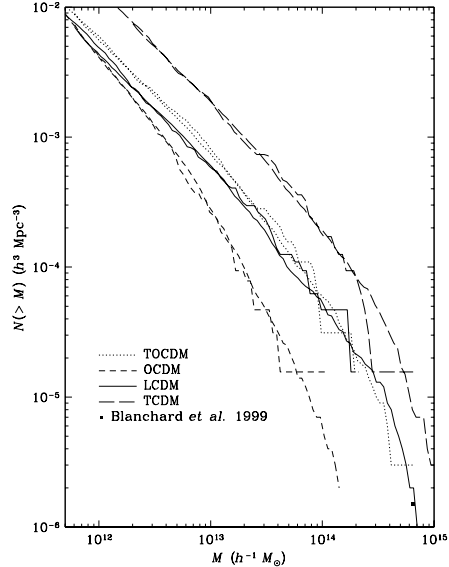,height=3in}
\caption{Halo Mass Functions}
\label{mf-fig}
\end{center}
\end{figure}

In \Fig{cmb-fig} we compare the theoretical CMB angular power spectra
of the models simulated here with the recent BOOMERanG~\cite{BOO}
anisotropy measurements.  In \Fig{mf-fig} we present cumulative halo
mass functions for the two sets of models simulated. Our two
cluster-normalized models, LCDM and TOCDM, reproduce similar mass
functions. The observational point in the figure~\cite{bla99} $ n \, (T
> 4.0 {\mathrm{keV}}) = \scinum{1.5\pm0.4}{-6} h^3 \mathrm{Mpc}^{-3} $
is in good agreement with the LCDM cluster abundance. The TOCDM curve
follows closely the LCDM curve, but for the former cosmology the
observational point would be shifted along the horizontal axis by a
factor $0.3^{1/3}$. However, it should be noted that there are still
significant uncertainties associated with both the observational
measurements of cluster abundance and the theoretical modeling of the
$M$--$T$ relation.~\cite{vl99}

There are two curves for each cosmological model in \Fig{mf-fig}---one
representing the mass function as measured in the \hmpc{100} box, the
other measured in the \hmpc{40} box. As illustrated in the figure, for
each model there is excellent agreement between the two curves.

\section{Summary}
\label{summary}

In this paper we introduce two matching sets of four cosmological
models. We derive halo mass functions for all models and use the small
box, high resolution simulations in order to verify the validity of
the mass function in the large box for halos as small as $\approx
10^{12} \OhMsol$. The simulations presented here are unique as they
both cover a volume comparable to current large \tred{} redshift
surveys {\em and} at the same time resolve cluster masses down to
\Mstar.
While the simulations were designed mostly in order to achieve
\cobe{}---and, when possible (LCDM and TOCDM), also
cluster---normalization, they also serve to demonstrate the
attractiveness of the LCDM model. Requests for the simulations
presented in this paper, or for other data from the \textit{Texas
  \pppm{} Database}, should be sent to {\tt
  database@galileo.as.utexas.edu}.

\section*{Acknowledgments}
We are indebted to Mike Gross for his stimulating help and friendship.
We thank Ue-Li Pen for helpful discussions and comments. HE was
supported by a Smithsonian Predoctoral Fellowship.  This work was
supported by NASA grants NAG5-7363 and NAG5-7812; NSF grant ASC
9504046; and the Texas Advance Research Program grant 3658-0624-1999.


\section*{References}
\begin{thebibliography}{99}

\bibitem{bla99}A. Blanchard \etal, \aap, submitted, astro-ph/9908037 (1999).
  
\bibitem{copi95}C.J. Copi, D.N. Schramm and M.S. Turner,
  \Journal{\apj}{455}{95}{1995}.
  
\bibitem{BOO}P. de~Bernardis \etal{}, \Journal{\nat}{404}{955}{2000}. 

\bibitem{gross98}M.A.K. Gross, R.S. Somerville, J.R. Primack, J.
  Holtzman and A. Klypin, \Journal{\mnras}{301}{81}{1998}. 
  
\bibitem{mm00}H. Martel and R. Matzner, \Journal{\apj}{530}{525}{2000}. 
  
\bibitem{mme00}H. Martel, R. Matzner and H. El-Ad, in preparation (2000).
  
\bibitem{pen98}U.-L. Pen, \Journal{\apj}{498}{60}{1998}. 

\bibitem{riess98}A.G. Riess \etal, \Journal{\aj}{116}{1009}{1998}. 
  
\bibitem{vl99}P.T.P. Viana and A.R. Liddle, \Journal{\mnras}{303}{535}{1999}. 

\end{thebibliography}
\end{document}